# White Dwarfs as Physics Laboratories: Lights and Shadows

J. Isern[1,2,3]*, S. Torres[2,4] and A. Rebassa-Mansergas[2,4]

[1]Institute for Space Sciences (ICE, CSIC), Cerdanyola del Vallès, Spain, [2]Institut d'Estudis Espacials de Catalunya (IEEC), Barcelona, Spain, [3]Fabra Observatory (RACAB), Barcelona, Spain, [4]Departament de Física, Universitat Politècnica de Catalunya (UPC), Barcelona, Spain

The evolution of white dwarfs is essentially a gravothermal process of cooling in which the basic ingredients for predicting their evolution are well identified, although not always well understood. There are two independent ways to test the cooling rate. One is the luminosity function of the white dwarf population, and another is the secular drift of the period of pulsation of those individuals that experience variations. Both scenarios are sensitive to the cooling or heating time scales, for which reason, the inclusion of any additional source or sink of energy will modify these properties and will allow to set bounds to these perturbations. These studies also require complete and statistical significant samples for which current large data surveys are providing an unprecedented wealth of information. In this paper we review how these techniques are applied to several cases like the secular drift of the Newton gravitational constant, neutrino magnetic moments, axions and weakly interacting massive particles (WIMPS).

## 1 INTRODUCTION

At present, the behavior of Nature is described with two fundamental theories. One is the General Relativity (GR), which describes the gravitational interaction, the other is the Standard Model (SM), which describes the electromagnetic, weak and strong interactions. According to Uzan (2011) these theories involve fields, symmetries and constants that are postulated in order to construct a mathematically consistent description of physical phenomena in the most unified and simple way. A fundamental constant of a physical theory is any parameter that cannot be explained by this theory. This means that these constants or combination of constants have to be measured with the maximum accuracy and precision. This is the case of the Newton constant, $G_N$, in gravitation for instance.

One of the main difficulties that the SM has to face is the existence of 20 parameters whose values are not predicted by the theory and have to be determined by experiments, and it has to face important challenges like accounting for the nature of the dark matter and dark energy, the matter/anti-matter asymmetry, the strong CP problem and so on. The difficulty is that, in many cases, the energies and/or the timescales involved in the determination of these parameters are so large that these experiments are beyond the possibilities of the present terrestrial laboratories.

On one hand, physical conditions covered by stars are very large, not only in densities and temperatures but also in energies and gravity strengths. These properties make them a useful complement of terrestrial laboratories. Obviously this is possible because there is a solid observational background[1] and, on the other hand, because at present stellar evolution is a predictive theory anchored over solid roots. Furthermore, the lifetime of many stars is of the order or larger than the age of the Universe for which reason they can be used as a reliable tool to check the evolution of the fundamental constants with time.

---

[1]Gaia mission is impressively improving the accuracy and precision of the stellar fundamental parameters.

Stars are poor photon emitters because the opacity of matter is so high that they can only escape from the layers above the photosphere. Therefore, the luminosity of stars is proportional to their surface, $S_{ph}$, and to the four power of the temperature of such a surface, $T_{eff}$, which is several orders of magnitude smaller than that of the interior, $L_{ph} \propto S_{ph} T_{eff}^4$. If stars were able to emit weakly (feebly) coupled low-mass particles (like neutrinos), the total luminosity, $L_{fc}$, would be proportional to the volume and to some power, $\alpha$, of the core temperature, $L_{fc} \propto V T_c^\alpha$, and at some stages of the evolution such a mechanism could be dominant (for instance, neutrinos are dominant beyond the helium-burning stage of the evolution of normal stars, or during the first stages of the cooling of white dwarfs and quite important during the red giant branch (RGB) evolution of low-mass stars). Therefore, stars can act as a powerful tool to obtain information about the feebly interacting low-mass particles predicted by high-energy field theories that are trying to go beyond the SM.

White dwarf stars have proved to be excellent laboratories for testing new physics since, on one hand, their evolution is, at a first order, a gravothermal process of cooling of a star in hydrostatic equilibrium and, on the other hand there is at present a solid observational background that allows a detailed comparison between the predictions of new physics and the observed data. Their use as laboratory tools is based on an energy-conservation argument (Raffelt, 1996). Traditionally, the argument was that since their evolution is a process of cooling, the minimum luminosity that an ensemble of white dwarfs can reach is determined by the age of the ensemble and the physical ingredients considered i.e., because of the finite age of the ensemble a lower limit of luminosity must exist. There are, however, two other independent ways to measure the cooling rate of WDs, one is based on the secular drift of the pulsation period of individual variables and the other on the shape of the luminosity function (Isern and García-Berro, 2008).

These procedures have allowed to put bounds on the mass of axions (Raffelt, 1986; Isern et al., 1992, 2008), on the neutrino magnetic momentum (Blinnikov and Dunina-Barkovskaya, 1994), the secular drift of the Newton gravitational constant (Vila, 1976; García-Berro et al., 1995), the density of magnetic monopoles (Freese, 1984) and WIMPS (Bertone and Fairbairn, 2008), as well as constraints on properties of extra dimensions (Malec and Biesiada, 2001)), on dark forces (Dreiner et al., 2013), on modified gravity (Saltas et al., 2018), and formation of black holes by high energy collisions (Giddings and Mangano, 2008). Only the cases of the drift of the Newton constant, the magnetic momentum of neutrinos, axions and WIMPs will be discussed in this review.

## 2 THE EVOLUTION OF SINGLE WHITE DWARFS

White dwarfs are the final evolutionary stage of low- and intermediate-mass stars ($M < 8$–$10$ M$_\odot$). They have a relatively simple structure composed by a degenerate core that contains the bulk of mass and acts as a reservoir of energy, and a partially degenerate envelope that controls the energy outflow. This envelope is formed by a thin helium layer with a mass of the order of ℏ $10^{-2}$ M$_\odot$, which, in turn, is surrounded by an even thinner layer of hydrogen with a mass in the range of $10^{-15}$—$10^{-4}$ M$_\odot$, although about 20% of white dwarfs do not have such hydrogen envelopes. White dwarfs displaying hydrogen in their spectra are known as DA and the remaining ones as non-DA[2].

The mass distribution of single DA white dwarfs is strongly peaked around 0.59 M$_\odot$ with clear indications of contamination by merged white dwarfs (Kilic et al., 2020) and a similar behavior, although with differences appears in the DB distribution (Koester and Kepler, 2015; Rolland et al., 2018; Tremblay et al., 2019) Those with a mass larger than $\sim 1.05$ M$_\odot$ have a core made of O and Ne. The remaining ones, the vast majority, have a core made of a mixture of C and O.

As a consequence of the high electronic degeneracy of their core, white dwarfs cannot obtain energy from nuclear reactions and their evolution is just a process of contraction and cooling. See Lamb and van Horn (1975); Iben and Tutukov (1984); Koester and Schoenberner (1986); D'Antona and Mazzitelli (1989); Isern et al., 1998; Fontaine et al., 2001; Hansen (2004); Althaus et al., 2010a; García-Berro and Oswalt (2016) for detailed descriptions of the cooling process.

This cooling process can be roughly divided into four stages (see **Figure 1**): neutrino cooling (log ($L/L_\odot$) >—1.5), fluid cooling (—1.5 ≥ log ($L/L_\odot$) ≥—3), crystallization (log ($L/L_\odot$) ≤—3) and Debye cooling. The energy balance of a white dwarf can be represented by

$$L + L_\nu = -\int_{M_{WD}} c_V \frac{dT_C}{dt} dm - \int_{M_{WD}} T\left(\frac{\partial P}{\partial T}\right)_{V,x} \frac{dV}{dt} dm + g_s + (l_s + e_s)\dot{m}_s \pm (\dot{\varepsilon}_e) \quad (1)$$

where the left hand side contains the energy losses (photons and neutrinos), while the right hand side contains the two terms associated to the gravo-thermal readjustment of the structure plus the gravitational settling of heavy species like $^{22}$Ne, $g_s$, the latent heat, $l_s$, and sedimentation associated to crystallization, $e_s$, times the crystallization rate $\dot{m}_s$ and any other exotic source or sink of energy, $\dot{\varepsilon}_e$. This equation has to be complemented with a relationship connecting the temperature of the core with the luminosity of the star. Typically $L \propto T_c^\beta$ with $\beta \approx 2.5$–2.7.

There are, however, several persisting problems. One of them is to determine the initial conditions under which white dwarfs form since many thermal structures can coexist during these early stages. In fact, these initial conditions are very complex and strongly dependent on the amount of residual hydrogen left by the parent Asymptotic Giant Branch (AGB) star. If it is large enough, $M_H \geq 10^{-4}$ M$_\odot$, pp reactions never stop and they even become dominant at low luminosities.

---

[2]The non-DA group is composed by stars with different spectroscopic properties named, in order of decreasing temperatures, DO, DB, DC, DQ.

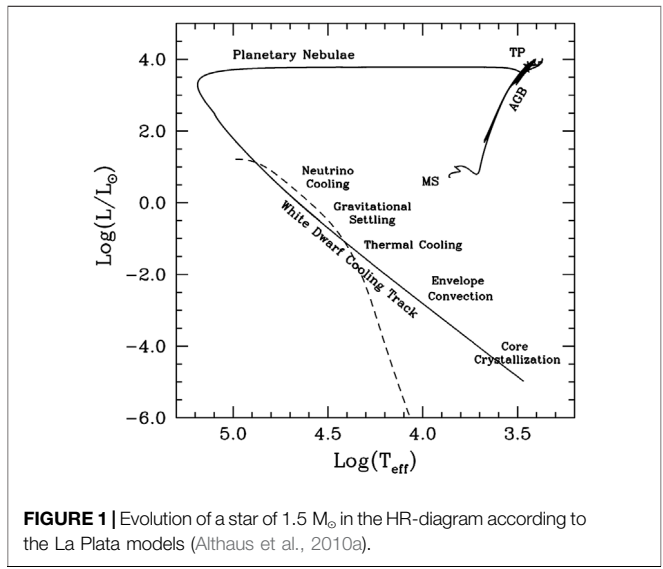

FIGURE 1 | Evolution of a star of 1.5 M$_\odot$ in the HR-diagram according to the La Plata models (Althaus et al., 2010a).

Fortunately, astero-seismological observations seem to constrain the mass of hydrogen well below this critical value and the dominant neutrino emission, forces the thermal structures produced by AGB stars to converge towards a unique one, guaranteeing in this way the uniformity of models dimmer than log $(L/L_\odot)$ =—1.5. Assuming exactly the same input physics, composition and stratification, the cooling times obtained by the LPCODE and BaSTI codes differ an 8% at this epoch just as a consequence of the different thermal structures of the initial converged models at the beginning of the cooling sequence (Salaris et al., 2013). This means that hot white dwarfs have to be used with care for testing new physics[3].

A related problem comes from the evolution of the envelope as the luminosity not only depends on the total mass and radius of the white dwarf, but also on the properties (mass, chemical composition and structure) of the outer layers. The main characteristics of the envelope is its tendency to become stratified, the lightest elements tending to be placed on top of the heaviest ones as a consequence of the strong gravitational field. However, this behavior is counter balanced by convection, molecular diffusion and other processes that tend to restore the chemical homogeneity. In any case, the ~80% of white dwarfs shows the presence of H–lines in their spectra (the DAs) while the remaining ~20% not (the non-DAs). This proportion is not constant along the cooling sequence. The most common interpretation is that the DAs have a double layered envelope made of H ($M_H \leq 10^{-4} M_\odot$) and He ($M_{He} \sim 10^{-2} M_\odot$) while the non-DAs have just a single He layer or an extremely thin H layer. An additional complication is that the initial conditions at the moment of formation are not well known and for the moment it is not possible to disentangle which part of this behavior is inherited and which part is evolutionary, although probably both are playing a role (Althaus et al., 2010a). In any case it is possible to adjust the parameters of the AGB progenitors to obtain 20% of white dwarfs completely devoid of the hydrogen layer. But, since the relative number of DA/non-DA stars changes during their evolution, a mechanism able to transform this character must exist (Shipman, 1997).

It is commonly accepted that DAs start as the central star of a planetary nebula with a hydrogen layer of a mass in the range of $10^{-8} - 10^{-4}$ M$_\odot$. As they cool down, the outer convection zone deepens and, depending on the mass, the hydrogen layer is completely engulfed by the larger helium layer in such a way that DAs turn out into non-DAs and, consequently, the ratio DA/non-DA decreases with the effective temperature. The evolution of non-DAs is more complex. They are thought to be the descendants of the He-rich central stars of planetary nebulae and, initially, as they cool down they look as PG 1159 stars first and DO after. Meanwhile, the small amount of hydrogen present in the envelope floats up to the surface and when the temperature is ~50,000 K forms an outer layer thick enough to hide the helium layer to the point that the star becomes a DA. When the temperature goes below 30,000 K, the convective helium layer engulfs the hydrogen one and the white dwarf recovers the non-DA character, now as a DB, and, as it continues to cool down, it becomes a DC (notice that a fraction of DCs has a DA origin). The lack of non-DA white dwarfs in the temperature range 50,000–30,000 K is known as the DB gap. Besides the phenomenological differences between DA and non-DA families, the most important property is that they cool down at different rates since hydrogen is more opaque than helium.

The main source of energy during the fluid phase is the gravothermal one. Since the plasma is not very strongly coupled ($\Gamma < 179$), its properties are reasonably well known (Segretain et al., 1994; Potekhin and Chabrier, 2010; Jermyn et al., 2021). Furthermore, the flux of energy through the envelope is controlled by a thick non-degenerate layer with an opacity dominated by hydrogen (if present) and helium, and weakly dependent on the metal content. A key ingredient is the electron conductivity in the frontier between moderate and strong degeneracy. Cassisi et al., 2021 and Salaris et al., 2022 have computed two sets of models using the electron conductivities obtained by Cassisi et al., 2007 and by Blouin et al., 2020b and have found that models computed with the Blouin et al. conductivities were predicting longer cooling times at bright luminosities and shorter cooling times at fainter luminosities as compared with the Cassisi et al. conductivities.

One source of uncertainty is related to the chemical structure of the interior, which depends on the adopted rate of the $^{12}$C $(\alpha,\gamma)$ $^{16}$O reaction and on the treatment given to semiconvection and overshooting (Salaris and Cassisi, 2017). If this rate is high, the oxygen abundance is higher in the center than in the outer layers, resulting in a reduction of the specific heat at the central layers where the oxygen abundance can be as high as $X_O = 0.85$ (Salaris et al., 1997). Fortunately, asteroseismological techniques have started to provide information about the internal chemical

---

[3]Chen et al., 2021 have reanalyzed the two twin galactic globular clusters M3 and M13 and have found a clear overabundance of bright WDs in M13. They interpret this difference as caused by the residual thermonuclear burning of hydrogen in the outer layers of low metallicity white dwarfs.

structure (De Gerónimo et al., 2017; Romero et al., 2017; Giammichele et al., 2018).

An additional source of uncertainty is the role of minor species like $^{22}$Ne. This isotope is the result of the α-burning of the $^{14}$N left by the hydrogen burning stage and its abundance is of the order of the sum of the carbon, nitrogen and oxygen abundances, $X^{22}Ne ≈ 0.02$ for solar metallicities. Because of its neutron excess and the high sensitivity of degenerate stellar structures on the electron number profile, $Y_e$, its migration to the central regions can represent an important source of gravitational energy despite its low abundance (Isern et al., 1991). A similar role can be played by $^{56}$Fe (Xu and van Horn, 1992), and other species with small Z/A like $^{13}$C or $^{18}$O. Since the timescale of cooling is very large, of the order of several Gyr, it is possible to foresee two processes of chemical differentiation: 1) gravitational diffusion due to the low electron mole number of $^{22}$Ne as compared with the average (Bravo et al., 1992), and 2) sedimentation induced by a phase transition (Isern et al., 1991).

During the fluid phase a mixture of C/O/Ne remains mixed (Ogata et al., 1993) and the only way to induce a migration of $^{22}$Ne towards the central regions is through gravitational diffusion. The importance of this settling can be understood with the following argument: in the case of a degenerate structure in equilibrium, the upwards electric field acting over a proton is $eE ≈ 2m_p g$, where $g = Gm(r)/r^2$ is the gravity induced by the inner mass $m(r)$, while the downwards force acting on a $^{22}$Ne nucleus is $F = 22m_p g − 10eE = 2m_p g$ (Bravo et al., 1992; Bildsten and Hall, 2001). Integrating this expression over the total radius of the star gives an estimation of the potential energy of 250 and 1,600 keV per nucleus, respectively, in the case of a 0.6 and a 1.2 M$_⊙$ white dwarf. Adopting the treatment of Hansen (1978) and Hameury et al., 1983, the corresponding local diffusion timescale was estimated to be (Bravo et al., 1992),

$$\tau (Gyr) = 2.24 T_8^{-1/3} m^{-1} \rho_8^{11/18} \left[ Z \left| \frac{\langle A \rangle}{\langle Z \rangle} - \frac{A}{Z} \right| \right]^{-1} \quad (2)$$

where $T_8$ and $\rho_8$ are the temperature and density, respectively, in units of $10^8$, which suggests that diffusion is efficient only in the envelope and in hot interiors, $T_8 ≥ 1$. However, since this early epoch lasts for a short time as a consequence of the neutrino emission, they concluded that $^{22}$Ne was not able to appreciably migrate before crystallization unless the diffusion coefficient was improved by an order of magnitude. This question was reexamined by Bildsten and Hall (2001) and Deloye and Bildsten (2002) with an improved physics input and concluded that, effectively, this mechanism was not efficient in low mass white dwarfs, the majority, but it could work in massive ones, $M \gtrsim 1.0$ M$_⊙$.

Selfconsistent calculations assuming that the diffusion within the crystal was negligible have shown that $^{22}$Ne is only depleted in the outermost layers and that this mechanism was only effective in massive white dwarfs[4] (García-Berro et al., 2008; Althaus et al., 2010b; Camisassa et al., 2016), in agreement with the early guess of Bravo et al., (1992). Similar results have been obtained by Bauer et al., (2020), Camisassa et al., (2021) and Salaris et al., (2022). Bauer et al., (2020) also explored the possibility of formation of solid clusters of neon containing ≈ $10^3$ nuclei near the freezing point, but molecular dynamics simulations (Caplan et al., 2020) suggest that the abundance of neon is too small for such an effect.

---

[4]In the case of a 0.6 M$_⊙$ star the delay was found to be negligible, while in the case of a 1.06 M$_⊙$ was ≈ 3.2 Gyr in the case of pure carbon and only ≈ 0.6 Gyr in the case of pure oxygen.

When the temperature is low enough, the plasma experiences a phase transition and crystallizes into a classical body-centered cubic crystal (bcc), the detailed structure being rather uncertain since the free energy of the different Coulomb crystals is very similar at low temperatures. In the case of one component plasma with atomic number Z, this occurs when the Coulomb parameter $\Gamma = Z^2 e^2 / a k_B T$, where $a = [3/(4\pi n_i)]^{1/3}$ is the ion-sphere radius, $n_i$ is the ion number density and $k_B$ is the Boltzmann constant, becomes larger then ≈ 175. Solidification introduces two additional sources of energy in the cooling process, latent heat and gravitational sedimentation.

The contribution of the latent heat was early recognized (van Horn, 1968; Shaviv and Kovetz, 1976). It is of the order of $k_B T_s$ per nuclei, where $k_B$ is the Boltzmann constant and $T_s$ is the temperature of solidification. Therefore, its contribution to the total luminosity is small, ≈ 5%, but not negligible.

During the process of crystallization of a C/O mixture the chemical composition of the solid and liquid plasmas that are in equilibrium are not equal. Therefore, if the resulting solid flakes are denser than the liquid mixture, they sink towards the central region. If they are lighter, they rise upwards and melt when the solidification temperature, which depends on the density as $T_S \propto \rho^{1/3}$, becomes equal to that of the isothermal core. Meanwhile, depending on their density, the liquid re-homogenizes via Rayleigh-Taylor instabilities. The net effect is a migration of the heavier elements towards the central regions and a release of gravitational energy (Schatzman, 1982; Mochkovitch, 1983; Isern et al., 1997a, 2000). Of course, the efficiency of the process depends on the detailed chemical composition and on the initial chemical profile and decreases if the abundance of oxygen is already higher in the central regions of the star.

The first to realize that Coulomb plasmas could experience a change of miscibility during the process of solidification was Schatzman (1958), but it has been necessary to wait for the detailed calculations of phase diagrams to explore the consequences of such an effect. The first phase diagram of the two dominant chemical species, $^{12}$C and $^{16}$O, was obtained by Stevenson (1980) who found an eutectic behavior with a complete separation of both species at the solid phase. These calculations were improved by Barrat et al., (1988) and Segretain and Chabrier (1993) who found a phase diagram of the spindle form (**Figure 2A**), while Ichimaru et al., 1988, Ogata et al., 1993, found an azeotropic behavior (**Figure 2B**) . In both cases, given the expected relative abundances of carbon and oxygen in white dwarfs, the solid that forms contains more oxygen and is denser than the liquid (see **Figures 2A,B**). Consequently, the solid settles down and the liquid with an excess of carbon that is left behind is redistributed by Rayleigh-Taylor instabilities. The result is an enrichment of oxygen in the central layers and its depletion in the outer ones, with the subsequent release of gravitational energy (Hernanz et al., 1994; Segretain et al., 1994; Isern et al., 1997b, 1998, 2000; Renedo et al., 2010; Salaris et al., 2010; Blouin et al., 2020a). The main difference between the spindle and the azeotropic behaviors is that in the second case the temperature of solidification is smaller, the energy is released at lower luminosities, and the delay introduced in the cooling is larger. This phase diagram has been reexamined by Horowitz et al., (2010), Medin and Cumming (2010) and Blouin et al., (2020a) who have confirmed the azeotropic behavior of the mixture upon crystallization.

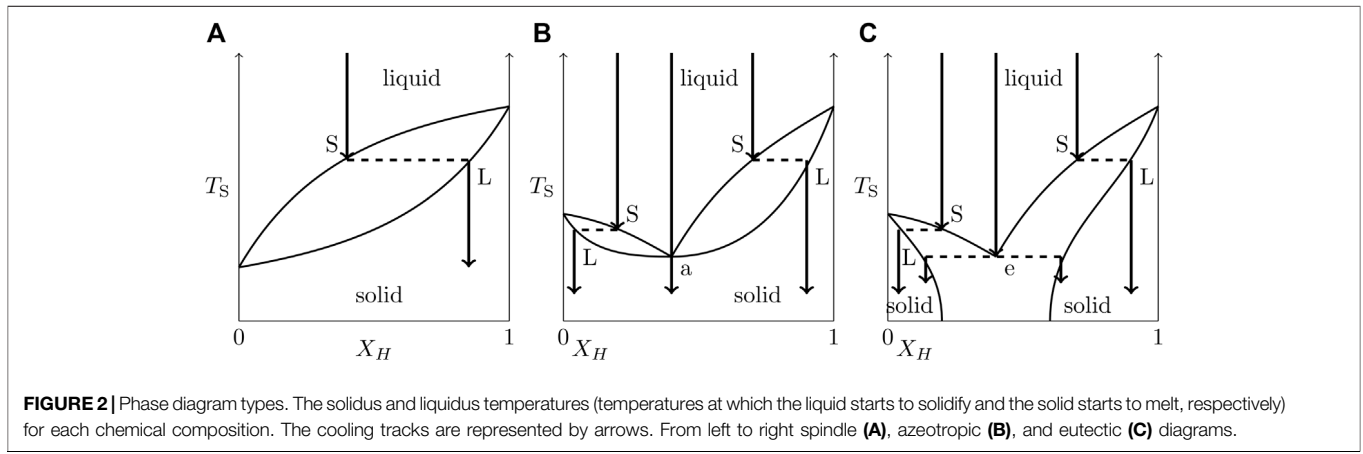

FIGURE 2 | Phase diagram types. The solidus and liquidus temperatures (temperatures at which the liquid starts to solidify and the solid starts to melt, respectively) for each chemical composition. The cooling tracks are represented by arrows. From left to right spindle (A), azeotropic (B), and eutectic (C) diagrams.

TABLE 1 | Energy release by crystallization and associated delays assuming a thick hydrogen envelope for a 0.6 $M_\odot$ white dwarf.

| Mixture | $\Delta E$ (erg) | $\Delta t$ (Gyr) |
| --- | --- | --- |
| C/O | 2.0, ×, $10^{46}$ | 1.8 |
| A/Ne | 1.5 × $10^{47}$ | 9 |
| A/Fe | 2.9, ×, $10^{46}$ | 1.1 |
| C/O/Ne | 0.2 × $10^{46}$ | 0.6 |

The delay introduced by solidification can be easily estimated to a good approximation if it is assumed that the luminosity of the white dwarf is just a function of the temperature of the nearly isothermal core (Isern et al., 1998). In this case

$$\Delta t = \int_0^{M_{WD}} \frac{\varepsilon_g(T_c)}{L(T_c)} dm \quad (3)$$

where $\varepsilon_g$ is the energy released per unit of crystallized mass and $T_c$ is the temperature of the core when the crystallization front is located at mass $m$. Table 1 displays the energy release and the corresponding delay associated to several chemical compositions. Of course, the total delay essentially depends on the transparency of the envelope. Any change in one sense or another can amplify or damp the influence of solidification. Table 1 displays an early calculation of the energy released in a 0.6 $M_\odot$ white dwarf made of half carbon and half oxygen (Isern et al., 1998).

The presence of neutron rich impurities, like $^{22}$Ne and $^{56}$Fe, is another source of uncertainty in the process of solidification. Isern et al., (1991) assumed it was possible to approach the ternary mixture C/O/Ne by a binary one consisting of neon and an average with atomic number $<Z_A> = 6n_{12} + 8n_{16}$, being $n$ the number fraction. Under such hypothesis they found an azeotrope for abundances of neon in the range of 0.05–0.09 by mass. Since the amount of neon in white dwarfs is expected to be smaller than this amount, the track in the phase diagram (Figure 2B) shows that the solid is poorer in neon, thus lighter, than the liquid and floats outwards where it melts and mixes with the ambient. The final outcome is the formation of a core, with the azeotropic composition, containing all the neon of the star.

Under the same hypothesis of an average nucleus representative of the mixture Segretain et al., (1994) found an azeotrope for neon at $X_a = 0.13$ and an eutectic (Figure 2C) for iron at $X_e = 0.75$ by number fraction leading in both cases to the formation of a heavy core containing neon and iron with the azeotropic and the eutectic concentrations (Isern et al., 1998). Table 1 displays the energy release and the subsequent delay introduced by this effect (cases A/Ne and A/Fe, respectively).

The use of an average nucleus is probably justified in the case of impurities of very high atomic number such as iron. However, in the case of Ne this assumption is probably doubtful, as has been shown by Segretain (1996), who examined the behavior of the ternary mixture and found that if the abundance of oxygen is high, the presence of neon plays a minor role and the phase diagram is similar to a pure carbon-oxygen diagram. Since in the process of sedimentation the liquid is progressively enriched in carbon and depleted in oxygen, there is a critical point, an azeotrope, with a neon concentration $X_{Ne} = 0.22$, a carbon concentration $X_C = 0.78$ and a null oxygen concentration $X_O = 0$, so that during the solidification process, the fluid crystallizes forming a shell made of a pure neon-carbon mixture. From the energetic point of view the energy released is much smaller than in the case A/Ne (Table 1). However, Caplan et al., (2020), using molecular dynamics techniques, concluded that neon cannot separate for realistic abundances[5] while Blouin et al., (2021), using a Clapeyron integration technique concluded that neon can separate and, depending on the chemical profiles, it was possible to obtain a neon-rich core or a carbon-neon shell without oxygen.

---

[5]In fact, their phase diagram allows to form a solid depleted of neon when the abundance of $^{22}$Ne is small allowing the formation of a neon-rich core.

It is interesting to notice here that Gaia Collaboration et al., (2018) found in the domain of the color-magnitude diagram corresponding to white dwarfs three overdensities, the A, B, and Q branches. The first two may be acquainted by the standard DA and non-DA evolution, but the third one is not yet fully understood. Cheng et al., (2019) noticed that this overdensity is placed near the crystallization point of $M \gtrsim 1 \, M_\odot$ and it cannot be accounted with the standard C/O model suggesting the gravitational settling of $^{22}$Ne as the extra source of energy necessary to introduce a substantial delay in the cooling of ⩾ 8 Gyr. Furthermore, Bauer et al., 2020 have shown that this branch is consistent with the solidification of C/O cores but not with the O/Ne ones, and Camisassa et al., (2021) that this cores must be massive and have neon abundances as large as $X^{22}$]Ne) ⩾ 0.06 to obtain the required effects in contradiction with the evolution of single stars. On the contrary, if the phase diagram of Blouin et al., (2021) is correct it would be possible to explain the origin of the Q-branch and, perhaps, to open the possibility of relating the meteoritic neon-E anomaly with the collision of two white dwarfs in which, at least one of them, has a pure C/Ne shell (Isern and Bravo, 2018a,b).

Concerning the sedimentation of $^{56}$Fe, (Caplan et al., 2021), have recently examined the question of its sedimentation and have found that for low mass white dwarfs and solar abundances the delay introduced is ⩾ 1 Gyr, but it reduces to ⩾ 0.1 Gyr in the case of more massive white dwarfs, thus being unable to account for the existence of the Q-branch. Similar results have been obtained by Salaris et al., (2022) but there are still substantial uncertainties in the details of the process.

Finally, when almost all the star has solidified, the specific heat follows the Debye's law and the star cools down very quickly. However, the outer layers are still far from the Debye regime and they prevent, at least in the case of thick envelopes, the sudden disappearance of the white dwarf (D'Antona and Mazzitelli, 1989).

## 3 THE LUMINOSITY FUNCTION

The luminosity function of an ensemble of white dwarfs (WDLF) is defined as the number of white dwarfs of a given luminosity per unit of magnitude interval, i.e. the luminosity distribution of these stars. If it is assumed that white dwarfs are not destroyed and that the ensemble is closed, the number of white dwarfs is then:

$$n(l) \propto \int_{M_i}^{M_s} \Phi(M) \Psi(\tau) \tau_{\rm cool}(l, M) \, dM \qquad (4)$$

where

$$\tau = T - t_{\rm cool}(l, M) - t_{\rm PS}(M)) \qquad (5)$$

and $l = -\log(L/L_\odot)$ is minus the logarithm of the luminosity in solar units, $M$ is the mass of the parent star (for convenience all white dwarfs are labeled with the mass of the main sequence progenitor), $t_{\rm cool}$ is the cooling time down to luminosity $l$, $\tau_{\rm cool} = dt/dM_{\rm bol}$ is the characteristic cooling time, $M_s$ and $M_i$ are, respectively, the maximum and the minimum masses of the main sequence stars able to produce a white dwarf of luminosity $l$, $t_{\rm PS}$ is the lifetime of the progenitor of the white dwarf, and $T$ is the age of the population under study. The remaining quantities, the initial mass function, $\Phi(M)$ or IMF, and the star formation rate, $\Psi(t)$ or SFR depend on the astronomical properties of the stellar ensemble. Hidden in the equation there is a relationship connecting the mass of the progenitor with the mass of the white dwarf, the initial-final mass relationship or IFMR (Catalán et al., 2008a).

The first luminosity function was derived more than 5 decades ago by Weidemann (1968), and since then it has been noticeably improved thanks to the work of many authors–see García-Berro and Oswalt (2016) for a recent review. **Figure 2A** shows the state of the art at the end of the ninetees, when the samples contained few hundreds of stars. The monotonic behavior of this function clearly proves that the evolution of white dwarfs is a simple gravothermal process of cooling, while the sharp cut-off at low luminosities is the consequence of the finite age of the Galaxy (Winget et al., 1987). The position of this cut-off depends on the energy sources and sinks that have been adopted to compute the evolution of white dwarfs and it can be used to constrain the existence of any additional energy term. However, the poorly determined position of the cut-off as a consequence of the low-luminosity of stars in this region and the difficulty to distinguish between DAs and non-DAs, together with the poor understanding of the physical properties of such cool stars prevent for the moment to obtain anything better than a rough bound.

Given the nature of the problem, there is always a degeneracy between the population properties (SFR, IMF) and the adopted stellar models. The usual approach has been to assume that stellar models and IMF are enough well known that it is possible to determine the galactic SFR. Thus, the question is how to obtain information about new physics from this equation without knowing the SFR.

**Eq. 4** can be rewritten as:

$$n(l) \propto \langle \tau_{\rm cool} \rangle \int_{M_i}^{M_{\rm max}} \Phi(M) \Psi(T - t_{\rm cool} - t_{\rm PS}) dM \qquad (6)$$

If this equation is restricted to the bright white dwarfs, i.e. those for which $t_{\rm cool}$ is small, the lower limit of the integral is satisfied by low-mass stars and is almost independent of the luminosity. Thus, the shape of the luminosity function only depends on the averaged physical properties of white dwarfs as far as the bright population is dominated by low mass stars, i.e. those coming from Main Sequence stars with $M \lesssim 1 \, M_\odot$, if the mass spectrum of white dwarfs is not strongly perturbed by the adopted SFR (Isern and García-Berro, 2008; Isern et al., 2009).

The empirical WDLFs independently obtained from the large cosmological surveys, like the Sloan Digital Sky Survey (SDSS) (Harris et al., 2006; DeGennaro et al., 2008; Rebassa-Mansergas et al., 2015b; Munn et al., 2017), the Super COSMOS Sky Survey (SCSS) (Rowell and Hambly, 2011), the LSS-GAC survey (Rebassa-Mansergas et al., 2015a) and PanSTARS (Lam et al., 2019) have increased the sample to several thousands white dwarfs improving in this way the precision and accuracy of the luminosity function to the point that it has been possible to compare the observational and theoretical shapes. **Figure 3B** displays the WDLFs obtained with the SDSS and SCSS catalogues normalized to the $M_{bol} \approx 12$ bin. As it can be seen they almost coincide over a large part of the luminosity interval. At high luminosities, $M_{bol} < 6$, both functions display a large dispersion as a consequence of the fact that the proper motion method is not appropriate there. This problem has been solved by Krzesinski et al., (2009) using the UV-excess technique. For illustrative purposes **Figures 3C,D** display three star formation rates and the corresponding luminosity functions obtained with them. As it can be seen, regardless of the adopted SFR these functions coincide in the range $6-7 \lesssim M_{bol} \lesssim 13-14$ and in all cases the change in the slope is a consequence of the transition from neutrino to photon cooling.

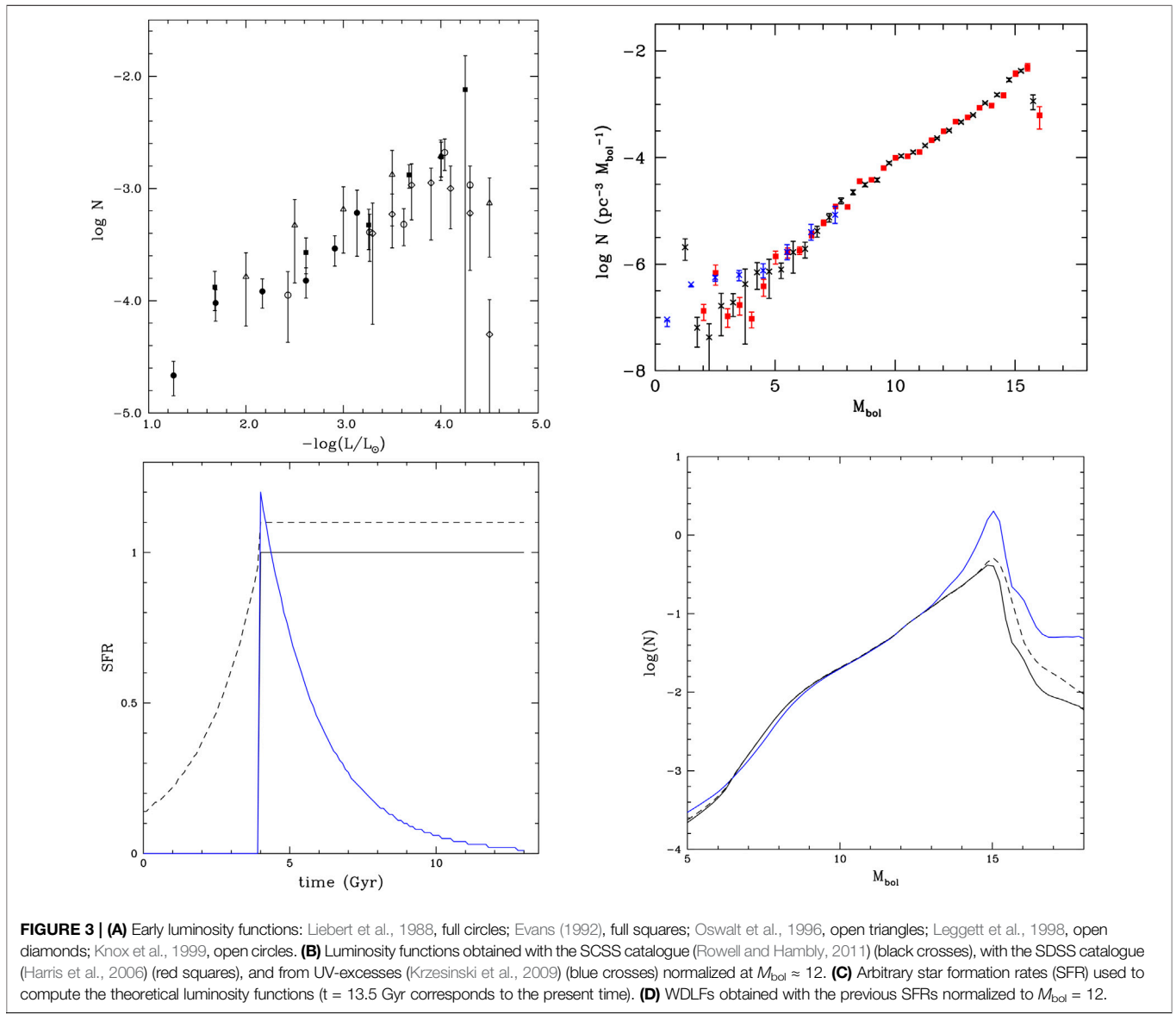

**FIGURE 3 | (A)** Early luminosity functions: Liebert et al., 1988, full circles; Evans (1992), full squares; Oswalt et al., 1996, open triangles; Leggett et al., 1998, open diamonds; Knox et al., 1999, open circles. **(B)** Luminosity functions obtained with the SCSS catalogue (Rowell and Hambly, 2011) (black crosses), with the SDSS catalogue (Harris et al., 2006) (red squares), and from UV-excesses (Krzesinski et al., 2009) (blue crosses) normalized at $M_{bol} \approx 12$. **(C)** Arbitrary star formation rates (SFR) used to compute the theoretical luminosity functions (t = 13.5 Gyr corresponds to the present time). **(D)** WDLFs obtained with the previous SFRs normalized to $M_{bol} = 12$.

The SFR functions previously considered are smooth and constant or declining with time. The existence of recent bursts of star formation can modify the dependence on the luminosity of the lower limit of the integral in **Eq. 6** and, adjusting bin by bin, it is possible to find, with models obtained with different physical ingredients, SFRs that match the observed luminosity function (Rowell, 2013). One way to break this degeneracy and to decide if it is necessary to introduce new physical effects is to examine the luminosity function of populations that have independent or poorly correlated star formation histories like the inner halo, and the thick and thin discs (Rowell and Hambly, 2011; Munn et al., 2017; Lam et al., 2019) and globular clusters (Bedin et al., 2008a; Hansen et al., 2015; Goldsbury et al., 2016) since any intrinsic effect has to appear in all of them (Isern et al., 2018).

The IMF plays a role of statistical weight in the definition of the WDLF and, at present, is the object of strong interest. It represents the mass distribution of stars of a single generation and it is found to be surprisingly universal in the solar neighborhood and beyond. Concerning its influence on the slope of the luminosity function it has been found it is minor as far as the number of stars born with a mass in the range of 0.8–1.0 $M_\odot$ is not strongly perturbed. For simplicity only the Salpeter function has been used here.

The IFMR links the mass of the white dwarf with that of its progenitor. The evolution during the Main Sequence is reasonably well known but the mass lost during the late stages, mainly during the AGB phase, remains elusive and prevents building a robust relationship between the two masses (Dominguez et al., 1999; Weiss and Ferguson, 2009; Renedo et al., 2010).

The empirical determination of this relationship is a hard task since the progenitor no longer exists. The method consists on obtaining the cooling age of the white dwarf and the total age of the star (white dwarf plus progenitor). Thus, the lifetime of the progenitor is the age of the star minus the cooling time and, if the metallicity is known, it is possible to obtain the mass via a mass-lifetime relationship (Sweeney, 1976). As it has be seen, for a given luminosity the age of the white dwarf depends on the mass, which requires accurate spectra, and the metallicity of the progenitor which is not known in the case of single ones.

One way to overcome these problems is to study white dwarfs in coeval systems like open clusters[6] (Kalirai et al., 2005; Dobbie et al., 2006; Kalirai et al., 2007; Cummings et al., 2018; Richer et al., 2021). The weak point in this case is that in order to compute the mass of white dwarfs it is necessary to obtain accurate spectra and this is only possible in the case of nearby open clusters. Furthermore, since these systems are young, white dwarfs are still hot and their age is uncertain (**Section 2**). Additionally, the range of metallicities covered is nearly solar.

Another way to obtain the IFMR is provided by white dwarfs in binaries that are wide enough to guarantee that stars have not interacted during their evolution. One possibility is the case of a white dwarf plus a main sequence star (Catalán et al., 2008a,b; Zhao et al., 2012) or a turn off/subgiant star (Barrientos and Chanamé, 2021). This procedure has the advantage that the metallicity of the system can be easily determined, with a plus, in the second case, that the total age of the system can be well evaluated. A potential problem is that, if the process of crystallization has already started, the effective temperature and the luminosity remain almost constant during the process and the cooling time is hard to obtain.

An alternative consists on using wide double white dwarfs. Since they are coeval and evolve idependently, the difference in the age of the progenitors is equal to the difference in the cooling ages. With these data it is possible to define a parametric IFMR and to find the set of parameters that best adjust the data (Andrews et al., 2015). As before, the problem is that the metallicity is not known and the ages of hot white dwarfs and crystallizing white dwarfs are uncertain.

Field WDs can also constrain the IFMR from their distribution in the colour-magnitude diagram since it depends on their cooling time and mass. El-Badry et al., (2018), using a sample of more than one thousand bright white dwarfs within a distance of ⨅ 100 pc were able to obtain a parameterized IFMR supporting the theoretically predicted non-linearity of this function (Salaris et al., 2009). As before, the metallicity is an important source of uncertainties since it can induce pile-ups in the regions of crystallization (Garcia-Berro et al., 1988; Hernanz et al., 1994; Kilic et al., 2020), making obvious the necessity of discussing the properties of the white dwarf population within a model of Galactic evolution.

Another source of uncertainty comes from the fact that an important fraction of stars are members of binary or multiple systems. De Rosa et al., (2014) have shown that among A-type main sequence stars 56% are single, 32% are in binaries and the remaining ones are in triple or multiple systems and, depending on the mass and separation, the resulting evolution can be completely different from that of isolated white dwarfs. This translates into modifications on the computed luminosity function and mass distribution of white dwarfs as compared with the single case. Furthermore, binaries can introduce important biases in the observed luminosity function (Rebassa-Mansergas et al., 2020).

With the only purpose to visualize the influence of mergers we have constructed a toy model that only considers situations in which the Roche lobe overflow occurs when the envelope of the star is convective and when magnetic braking and gravitational wave emission are effective (Isern et al., 1997a, 2013). Briefly, the model assumes that all the orbital energy is invested in evaporating the common envelope and that: 1) the IMF is written as the mass distribution of the primary times the mass ratio distribution, while the mass distribution of the primary star is a simple Salpeter's distribution in the range $0.1 \leq M_1/M_\odot \leq 100$ and the mass ratio distribution as $f(q) \propto q^n$, where $q = M_2/M_1$, where $M_1$ and $M_2$ are the masses of the primary and secondary, respectively, and $n = 1$, 2) the adopted distribution of separations is $H(A_0) = (1/5) \log(R_\odot/A_0)$, where $A_0$ is the initial separation and, in order to maximize the impact of mergers in the mass distribution function, it has been assumed that single white dwarfs are well represented by the distribution of wide binaries, 3) it is also assumed a constant star formation rate that started 9 Gyr ago, 4) the influence of metallicity on the age of the progenitors and on the mass of the resulting white dwarf has been neglected, and it has also been assumed that all white dwarfs more massive than 1.05 $M_\odot$ are made of oxygen and neon, 5) It has also been assumed that all binary systems are resolvable in mass, 6) the stellar data has been obtained from the BaSTI models (Salaris et al., 2010), and 6) C/O white dwarfs with $M_{WD} \geq 0.8\ M_\odot$ explode as sub-Chandrasekhar SNIa when they merge with a He-white dwarf. **Figure 3** displays the mass distribution for a single, a binary, and a mixed single-binary WD population obtained in this way. Temmink et al., (2020), using a more complete model, also reached the conclusion that mergers can modify the expected mass and luminosity distributions.

The *Gaia* mission has provided extremely accurate astrometric and photometric data of the local white dwarf population within (Gaia Collaboration et al., 2018) that has allowed to noticeably enlarge the number of identified WDs in the solar neighbourhood (Jiménez-Esteban et al., 2018; Gentile Fusillo et al., 2019, 2021) providing in this way a well defined and homogeneous sample of ℋ 359, 000 stars with a noticeable degree of completitude, ℋ 96 % within 20–100 pc but with only ℋ 10 % with accurate spectroscopic data. Despite that, these data have allowed noticeable discoveries about the WD population.

---

[6]The first one to use open clusters to otain the IFMR was Sweeney (1976).

In principle, volume limited surveys can provide unbiased empirical luminosity and mass distributions. However, the internal structure of galaxies evolve with time as a consequence of their interaction with neighbours and the existence of asymmetries in their gravitational potential. Recently plenty of structures have been identified in the halo and the thick disc that have been associated with several mergers in the past. The existence of the thin disc suggests that the Galaxy has not experienced a strong perturbation during its lifetime, ℋ 9 Gyr (Reid, 2005). Nevertheless, the interaction with the spiral arms, molecular clouds, massive stars, induce radial migrations[7] that generate an inflow/outflow of stars with multiple chemical compositions around the solar neighborhood and invalidate **Eq. 4**.

## 4 PULSATING WHITE DWARFS

During their evolution stars experience episodes of variability that are characterized by the existence of a specific spectrum of frequencies that depend on the details of their structure, i.e. they are characteristic of each star.

There are several modes of pulsation. The radial ones maintain the spherical symmetry while the non-radial ones do not. The radial modes are the simplest oscillations that a star can sustain and are those experienced by Cepheids and RR Lyr stars. The non-radial modes are classified into spheroidal and toroidal. The spherical ones are divided into g-, f-, and p-modes according to the restoring force. In the first two cases is gravity while in the third one is the pressure gradient. They can be described by

$$\psi'_{klm}(r, \theta, \varphi, t) = \psi'(r) Y_l^m(\theta, \varphi) e^{i\omega_{klm} t} \quad (7)$$

where the prime indicates a small Eulerian perturbation of a given physical variable and $Y_l^m$ is the corresponding spherical harmonics.

In general, g-modes are characterized by low oscillation frequencies and by horizontal displacements, while p-modes have higher frequencies and radial displacements. The structure of the g-modes is governed by the Brunt-Väisälä (BV) frequency

$$N_{BV}^2 = \frac{g^2 \rho}{P} \frac{\chi_T}{\chi_\rho} (\nabla_{ad} - \nabla + B) \quad (8)$$

where B is the Ledoux term

$$B = -\frac{1}{\chi_T} \sum_{i=1}^{N-1} \chi_{x_i} \frac{d \ln x_i}{d \ln P} \quad (9)$$

and $\chi_T = (\partial \ln P / \partial \ln T)_{\rho, \{x_i\}}$; $\chi_\rho = (\partial \ln P / \partial \ln \rho)_{T, \{x_i\}}$; $\chi_{x_i} = (\partial \ln P / \partial \ln x_i)_{T, \rho, \{x_{j \neq i}\}}$ and $\nabla$ and $\nabla_{ad} = (\partial \ln T / \partial \ln P)_{ad, \{x_i\}}$ are the real and the adiabatic gradients, $x_i$ is the chemical composition, and $g$ the local gravity acceleration. If $N_{BV}^2 > 0$ the fluid is in equilibrium and $N_{BV}$ is the frequency of oscillation of a parcel of fluid. If $N_{BV}^2 < 0$ the fluid becomes unstable and convection appears. The critical frequency of p-modes are governed by the Lamb frequency.

$$L_l^2 = l(l+1) c_s^2 / r^2 \quad (10)$$

where $c_s$ is the adiabatic sound speed $c_s^2 = \Gamma_1 P / \rho$. This means that a sound wave travels a distance $\approx 2\pi r / l$ horizontally in a time $\approx 2\pi / L_l$.

Within the Cowling approximation[8] the high radial order modes satisfy the following dispersion relation:

$$k_r^2 = \frac{1}{\omega^2 c_s^2} (\omega^2 - L_l^2)(\omega^2 - N_{BV}^2) \quad (11)$$

If $k_r^2 > 0$, the solution is oscillating. This happens when $\omega^2 > N_{BV}^2, L_l^2$ (high-frequency acoustic or p-mode oscillations) or when $\omega^2 < N_{BV}^2, L_l^2$ (low-frequency g-modes). These oscillations are within the zones limited by the condition $k_r^2 = 0$, which defines the turning points of the waves. When $N_{BV}^2 < \omega^2 < L_l^2$ or $N_{BV}^2 > \omega^2 > L_l^2$, $k_r^2 < 0$ and the solution evolves exponentially, generally as a decline except when there is a dynamical instability.

During their evolution, white dwarfs go across some regions of the H-R diagram where they become unstable and pulsate[9]. There are at least six classes of pulsating white dwarfs (Córsico et al., 2019), but here only DAVs, DBVs, and DOVs will be considered (**Table 2**). The multifrequency character and the amplitude of the period of pulsation ($10^2$–$10^3$ s) indicate they are g-mode pulsators, i.e. the driven mechanism is buoyancy. In principle, if there are enough data it is possible to obtain information about the mass, the internal chemical stratification, the rotation rate, the presence of magnetic fields, the cooling timescale and the core composition of the white dwarf. In this sense, the *Kepler* mission plus its *K2* extension (90 days of uninterrupted observations) have noticeably improved the number of variable white dwarfs that have been intensively studied and it is expected that the data from the missions *TESS*, *Cheops* and *Plato* will qualitatively change our knowledge of white dwarfs.

The start and quench of pulsations is not yet fully understood. Pulsations appear at an effective temperature in which the dominant chemical species become partially ionized (Dolez and Vauclair, 1981; Winget et al., 1982), is the so called $\kappa$—$\gamma$ mechanism, and/or when the outer convective envelope sinks (Brickhill, 1991; Goldreich and Wu, 1999), is the convective-driven mechanism. Both mechanisms predict reasonably well the effective temperature of the hot edge of the instability strip, but they fail in the cool boundary.

---

[7]Is the old problem of finding at which place the Sun was born (Wielen et al., 1996).

[8]The perturbations of the gravitational potential are negligible.

[9]See Althaus et al., 2010a; Córsico et al., 2019; Córsico (2020); Aerts (2021) for reviews on white dwarf asteroseismology.

TABLE 2 | Main characteristics of variable white dwarfs.

| Class | $T_{eff}$ | log g | Amplitude | Period | $\dot{P}$ |
|---|---|---|---|---|---|
| | K | (cgs) | (mag) | (s) | (s$^{-1}$) |
| DOV (GW Vir) | 80,000–180,000 | 5.5–7.7 | 0.02–0.1 | 300–2,600 | $10^{-10}$–$10^{-12}$ |
| DBV (V777 Her) | 22,400–32,000 | 7.5–8.3 | 0.05–0.3 | 120–1,080 | $10^{-12}$–$10^{-13}$ |
| DAV (ZZ Cet) | 10,400–12,400 | 7.5–9.1 | 0.01–0.3 | 100–1,400 | $(1-6) \times 10^{-15}$ |

As a consequence of the changes in the thermal and mechanical structure associated to the cooling process, the oscillation period, $P$, changes. Accurate modeling has shown that the introduction of additional distributed and moderate energy sinks or sources do not strongly modify the thermal profiles and, consequently, the spectrum of pulsations is not altered. However, this is not the case for secular drift rate, $\dot{P}$, which changes at a rate that can be approximated by (Winget et al., 1983)

$$\frac{\dot{P}}{P} = -a\frac{\dot{T}}{T} + b\frac{\dot{R}}{R} \quad (12)$$

where $a$ and $b$ are positive constants of the order of unity. The first term of the r.h.s reflects the decrease of the Brunt-Väisälä frequency with the temperature, while the second term reflects the increase of the frequency induced by the residual gravitational contraction. In the case of DOVs, gravitational contraction is still significant and the second term of **Eq. 12** is not negligible, for which reason $\dot{P}$ can be positive or negative, while the secular drifts of DBVs and DAVs are always positive and within the range of $10^{-12} - 10^{-13}$ ss$^{-1}$ and ≈ $10^{-15}$ ss$^{-1}$, respectively. Therefore, these secular drifts can be used to test the predicted evolution of white dwarfs and, if the models are reliable enough, to test any physical effect able to change the pulsation period of these stars. For instance, **Figure 5** displays the evolution of photon, neutrino and axion luminosities during the cooling process. It is clear that the introduction of an extra cooling mechanism will modify $\dot{P}$ and the extra luminosity approximately estimated as (Isern et al., 1992)

$$\frac{L_{ext} + L_{std}}{L_{std}} \approx \frac{\dot{P}_{obs}}{\dot{P}_{std}} \quad (13)$$

where *std* corresponds to the values obtained with the standard model, $L_{ext}$ is the extra source or sink, and $\dot{P}_{obs}$ is the observed period.

Notice that if there is a resonance between the local wavelength of a pulsation mode and the thickness of a layer, like the H or the He envelopes, the mode is trapped and the drift of the period can be substantially modified because the radial term of the r.h.s of **Eq. 12** is not negligible since these external layers are still contracting.

This method is based on the properties of individual stars and, consequently, less sensitive to the properties of the parent population. However, the drift not only depends on the cooling rate but also on the detailed structure of each individual. The uncertainties described in **Section 2** may have a deep impact on the calculation of $\dot{P}$. For instance changes in the estimated mass of the white dwarf or on its chemical profiles can introduce errors of ≈ 10% each one, while the thickness of the envelope layers can introduce, via mode trapping, changes of the order of 20–30% (De Gerónimo et al., 2017).

As in the case of the luminosity function, the presence of neutron rich impurities can introduce important modifications on the seismological properties of white dwarfs as a consequence of the strong dependence of the electronic pressure on the mean molecular weight per nucleon of the electrons and on the migration of neutron rich isotopes towards the center of the star. These modifications affect the values of the periods, their spacing and their evolution with time (see **Equation 9**) (Camisassa et al., 2016; Giammichele et al., 2018; Chidester et al., 2021).

The observational situation can be summarized as follows:

- DOVs or Pulsating PG 1159 or GW Vir stars. They are the hottest variables and the class contains white dwarfs with H-deficient, C/O/He-rich atmospheres and pre-white dwarfss. The first one was discovered by McGraw et al., (1979).

The pulsation drift has been measured in **PG1159-035** and amounts $\dot{P} = (1.52 \pm 0.05) \times 10^{-10}$ ss$^{-1}$ (Costa and Kepler, 2008).

- DBVs or V777 Her stars. Their atmosphere is made of almost pure helium. Their existence was predicted and discovered by (Winget et al., 1982). The total number known at present is 21 but the secular drift has only been measured in one of them:

**PG351 + 489** with $\dot{P} = (2.0 \pm 0.9) \times 10^{-13}$ ss$^{-1}$ for the largest amplitude period of ≈ 489 s (Redaelli et al., 2011). The expected theoretical drift is $\dot{P} = (0.81 \pm 0.5) \times 10^{-13}$ ss$^{-1}$ (Córsico et al., 2012b).

- DAVs or ZZ Ceti stars. They were the first to be discovered (Landolt, 1968) and are the most numerous. Their atmosphere is made of almost pure hydrogen, and they are characterized by low effective temperatures and high gravities. The total number of DAVs known at present is approaching to 300 (Córsico et al., 2019; Vincent et al., 2020), but the period drift has only been measured in three of them:

**G117-B15A** The monitoring of this star started in 1974 and is still continuing. Kepler (1984) demonstrated that the observed variability was due to non-radial g-modes, and Kepler et al., 2000 that the contribution of the still ongoing contraction was an order of magnitude smaller than that of the cooling. **Figure 6** displays the historical evolution of the measurement of the drift of the ̄H 215 s period mode. The present value is (5.12 ± 0.82) × $10^{-15}$ ss$^{-1}$ while the predicted one is (1.25 ± 0.09) × $10^{-15}$ ss$^{-1}$ (Kepler et al., 2021).

**TABLE 3** | Bounds to the variation of the Newton constant $G_N$.

| $\dot{G}_N/G_N$ (× $10^{-12}$ yr$^{-1}$) | Method | |
|---|---|---|
| 0.2 ± 0.7 | Lunar Laser Ranging | Hofmann et al., (2010) |
| ≤ 1.6 | Helioseismology | Guenther et al., (1998) |
| 0.05 ± 0.35 | Big Bang Nucleosynthesis | Copi et al., (2004); Bambi et al., (2005) |
| − 14 ± − 21 | Globular Clusters | degl'Innocenti et al., (1996) |
| − 0.6 ± 4.2 | Neutron stars in binaries | Thorsett (1996) |
| ≤ 20 | WDLF thin disk | Garcia-Berro et al., (2018) |
| − 0.9 ± 0.9 | NGC 6791 | García-Berro et al., (2011) |
| − 65 ± 65 | White dwarf variables | Córsico et al., (2013) |
| 6 ± 20 | Type Ia Supernovae | Lorén-Aguilar et al., (2003) |

**TABLE 4** | Bounds to neutrino magnetic-momentum.

| | $\mu_\nu/\mu_B \times 10^{12}$ | |
|---|---|---|
| Laboratory | < 29 | Beda et al., (2013) |
| Sun | < 400 | Raffelt (1999) |
| XENON1T | < 29 | Aprile et al., (2020) |
| Massive stars | < 20, −, 40 | Heger et al., (2009) |
| NGC 4258 | < 1.5 | Capozzi and Raffelt (2020) |
| WDV PG 1351489 | ≤ 7 | Córsico et al., (2014) |
| WDLF | < 5 | Miller Bertolami (2014) |
| WDLF 47 Tuc | < 3.4 | Hansen et al., (2015) |

**R548**, the ZZCeti star, has been monitored since 1970. The secular drift of its 213 s period is (3.3 ± 1.1) × $10^{-15}$ ss$^{-1}$ after subtracting the proper motion correction (Mukadam et al., 2013). This value is very similar to that obtained in G117-B15A which is not surprising since they have similar effective temperatures, masses and pulsation characteristics. The expected theoretical drift is (1.1 ± 0.09) × $10^{-15}$ ss$^{-1}$ (Córsico et al., 2012b).

**L19-2** is placed near the hot edge of the DAV instability strip. Its effective temperature is estimated to be $T_e \approx 12,100$ K and its mass ̄H 0.75 $M_\odot$ and has two pulsation modes not affected by trapping with periods 113 and 192 s periods respectively. The secular drift is (3.0 ± 0.6) × $10^{-15}$ ss$^{-1}$ in both cases (Sullivan and Chote, 2015), while the expected theoretical drifts are (1.42 ± 0.85) × $10^{-15}$ ss$^{-1}$ and (2.41 ± 1.45) × $10^{-15}$ ss$^{-1}$, respectively (Córsico et al., 2016).

It is clear that, given their smallness, the measurement of these secular drifts is extremely difficult and demands long observations. Up to now this has been done with the WET (World Earth Telescope), a large set of telescopes able to follow targets during long, nearly uninterrupted period of time (Nather et al., 1990). Fortunately *Kepler*, *TESS*, *Cheops* and *Plato* will soon alleviate the situation.

## 5 SECULAR DRIFT OF THE GRAVITATIONAL CONSTANT

The gravitation constant, $G_N$, plays a key role in the Theory of GR. In fact, as a consequence of the equivalence principle, $G_N$ must be a true constant, but this is just a hypothesis that must be verified. This constancy of the gravitational constant has been the object of a debate that started a century ago (Milne, 1935, 1937; Jordan, 1937) and that is still open. The main reason for such a long debate is twofold, one comes from the difficulties of measuring it, in fact, the gravitation constant is the worst measured one (Tiesinga et al., 2021), the other one is the formulation of alternative theories of gravitation in which $G_N$ is not constant but changes at a cosmological scale in space or time. This possibility, coupled with the key role that plays in the Theory of GR, has bursted the interest for detecting such variations or, at least, to put bounds as tight as possible. Consequently, there have been many attempts to measure a time variation of $G_N$ and several different methods have been proposed and used so far. **Table 3** displays some of the values obtained up to now. Most of these bounds come either from local measurements (the Sun and the Solar System or the local neighborhood) or from very early times (Big Bang nucleosynthesis, CMB) whereas at intermediate look–back times there are not so many measurements (Hubble diagram of SNIa). Here, only the bounds related with white dwarfs will be discussed.

Ranging methods are providing very precise and useful bounds but they are local limits and the same happens with the helioseismological ones. White dwarfs provide a not so precise but still useful local bound thanks to their luminosity function since they are a collectivity of objects with a large variety of ages and, at the same time, they contain some individuals, the variable ones, that admit a cross check of a hypothetical variability of $G_N$ via their secular pulsation drift. The first attempt to constrain $\dot{G}_N$ using the luminosity function was due to Vila (1976) but it was not successful because of the lack of reliable observational data and the uncertainties of the theoretical models.

When white dwarfs are cool enough their luminosity is entirely gravothermal. Changes in the value of $G_N$ translate into changes in the energy balance of the star and into the luminosity. This influence can be formally expressed as $L = -\dot{B} + \Omega(\dot{G}_N/G_N)$, where $B = U + \Omega$ is the total binding energy, being $U$ and $\Omega$ the total internal and gravitational energies respectively. Therefore if $\dot{G}_N \neq 0$ the luminosity and characteristic cooling time are different from those obtained in the case $\dot{G}_N = 0$ and, as it has been seen, this perturbation can modify the shape of the luminosity function and the position of the cut-off. It is important to remember here that the internal energy of white dwarfs is stored in the form of chemical potential (roughly the Fermi energy of electrons) and thermal kinetic energy for which reason, according to the virial theorem, only the second term contributes to the luminosity in the case of a varying $G_N$ (García-Berro et al., 1995).

**TABLE 5** | Bounds and hints to axion interactions ($g_{\gamma,10} = g_{a\gamma} \times 10^{10}$ GeV, $g_{e,13} = g_{ae} \times 10^{13}$).

| | Coupling constants | |
|---|---|---|
| Sun | $g_{\gamma 10} < 4.1$ | Vinyoles et al., (2015) |
| NGC 4258 (TRGB) | $g_{e,13} < 1.6$ | Capozzi and Raffelt (2020) |
| $\omega$ Cen (TRGB) | $g_{e,13} < 1.3$ | Capozzi and Raffelt (2020) |
| 22 Globular clusters | $g_{e,13} < 1.48$ | Straniero et al., (2020) |
| HB stars | $g_{\gamma 10} < 0.65$ | Ayala et al., (2014); Straniero et al., (2015) |
| G117-B15A | $g_{e,13} = 5.66 \pm 0.57$ | Kepler et al., (2021) |
| R548 | $g_{e,13} = 4.8 \pm 1.6$ | Córsico et al., (2012b) |
| L 19-2 (113) | $g_{e,13} = 4.2 \pm 2.8$ | Córsico et al., (2016) |
| L 19-2 (192) | $g_{e,13} < 5$ | Córsico et al., (2016) |
| PG 1351 + 489 | $g_{e,13} < 5.5$ | Battich et al., (2016) |
| WDLF | $g_{e,13} < 2.1$ | Miller Bertolami (2014) |

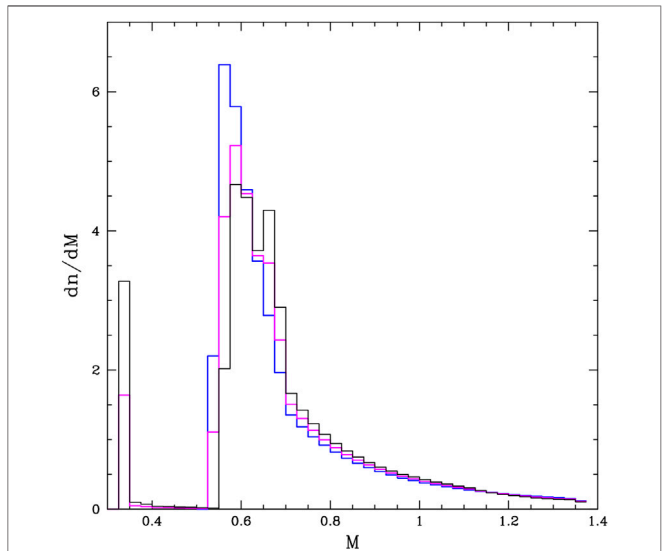

**FIGURE 4** | Normalized mass distribution of white dwarfs: single (blue), binaries (black), and a mixture obtained from fifty-fifty of astrated mass in form of binaries and singles.

An important point is that the relationship between the central temperature and the luminosity of the white dwarf also depends on $\dot{G}_N$ and for any source of energy, the cooling time delay is

$$\Delta t = \int_0^{M_{WD}} \frac{\varepsilon(T_c)}{L(G_N, T_c)} dm \quad (14)$$

Consequently, the cooling time is sensitive to the actual value of $G_N$ at each moment. In other words, the luminosity function is sensitive to the functional shape of the temporal variation of the gravitation constant (Althaus et al., 2011).

An additional improvement is the introduction of the dependence of the lifetime of the progenitors on the Newton constant, which can be obtained just from scaling arguments [degl'Innocenti et al., 1996]. For instance, in the case of 1 to 2 $M_\odot$ stars this lifetime can be approached by (García-Berro et al., 2011)

$$\tau_{MS} = \frac{1}{\gamma \left|\frac{\dot{G}_N}{G_N}\right|} \ln\left[\gamma \left|\frac{\dot{G}_N}{G_N}\right| \left(\frac{G_N^0}{G_N^i}\right)^\gamma \tau_{MS}^0 + 1\right] \quad (15)$$

where, $G_N^0$ and $G_N^i$ are the values of the Newton constant at present and when the star was born respectively and $\tau_M^0 S$ is the lifetime obtained with the present value of the gravitational constant. Once more, it is seen that the lifetime of the star depends on the functional form of $G_N(t)$.

Using the luminosity function of white dwarfs within 40 pc from the Sun (Limoges et al., 2015) and adopting an age of the Galactic thin disk of ≈ 9 Gyr, in agreement with the Th/Eu nucleocosmochronology (del Peloso et al., 2005), Garcia-Berro et al., (2018) obtained a bound of $\dot{G}_N/G_N \lesssim 2 \times 10^{-11}$ yr$^{-1}$ with cooling models including the gravitational diffusion of $^{22}$Ne and C/O separation upon crystallization.

This limit can be improved if the age of the population and the magnitude of the WDLF cut-off are known. NGC6791, for instance, is a metal rich open cluster with a turn-off age of $8.0 \pm 0.4$ Gyr. Its white dwarf luminosity function presents two peaks, the bright one corresponding to the population of unresolved binary white dwarfs and the faint one to the finite age of the cluster (Bedin et al., 2008a,b). When the diffusion and crystallization effects are taken into account the ages obtained from white dwarfs and from main-sequence stars coincide (García-Berro et al., 2010). Since the modulus of distance, obtained from eclipsing binaries, is $13.46 \pm 0.1$ (Grundahl et al., 2008) the luminosity of the cut-off is well determined. From both data it was obtained $|\dot{G}_N/G_N| \lesssim 1.8 \times 10^{-12}$ yr$^{-1}$ (García-Berro et al., 2011).

Another possibility is provided by the secular drift of g-mode pulsations (see **Eq. 12**. Since the restoring force driving pulsations is gravity, $\dot{T}$ and $\dot{P}$ depend on $\dot{G}_N$ (Benvenuto et al., 2004), the measure of the secular drift can also provide a useful constraint. The analysis of G117-B15A and R548 provided the bounds $\dot{G}_N/G_N = -1.79^{+0.53}_{-0.49} \times 10^{-10}$ yr$^{-1}$ and $\dot{G}_N/G_N = -1.29^{+0.71}_{-0.63} \times 10^{-10}$ yr$^{-1}$ respectively (Córsico et al., 2013). In this case only decreasing values of the gravitation constant where considered since the secular period drift of these stars is positive.

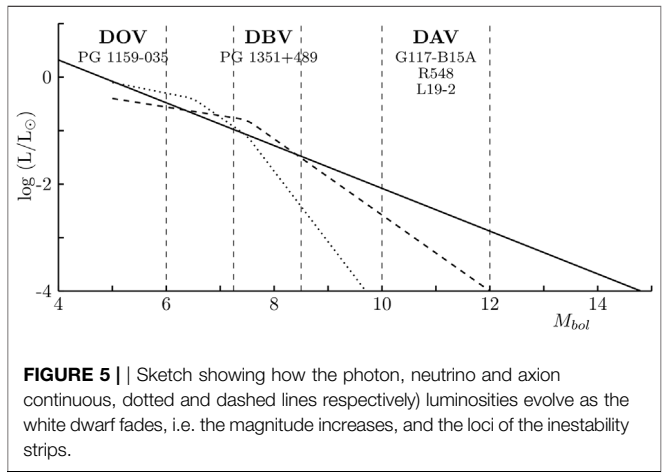

**FIGURE 5** | Sketch showing how the photon, neutrino and axion continuous, dotted and dashed lines respectively) luminosities evolve as the white dwarf fades, i.e. the magnitude increases, and the loci of the inestability strips.

The Chandrasekhar's mass, $M_{Ch} \propto G_N^{-3/2}$, also offers the possibility to obtain bounds on the variability of $G_N$. Although there are several subtypes of Type Ia supernovae it is thought that at least a fraction are the outcome of the explosion of a carbon-oxygen white dwarf near the Chandrasekhar's mass. In this case, the maximum of the light curve satisfies, $L_{max} \propto M_{Ni} \propto M_{Ch}$ (Arnett, 1982) and the evolution of this peak with redshift can be used to test the variation of $G_N$ with cosmic ages. Assuming that the fraction of mass converted into $^{56}N$ is a fixed fraction of the Chandrasekhar's mass, one obtains (Gaztañaga et al., 2002):

$$m_z = M_{bol,0} + \frac{15}{4}\log\left(G_N/G_{N,0}\right) + \ln d_L(z;\Omega_M,\Omega_\Lambda;G_N) + 25 \quad (16)$$

where $m_z$ is the apparent bolometric magnitude, $M_{bol,0}$ is the (intrinsic) absolute bolometric magnitude, 0 represents the present moment, and $d_L$ is the luminosity distance. Since the local bounds are $\dot{G}_{N,0}/G_{N,0} \lesssim 1 \times 10^{-12}$ yr$^{-1}$ it would be necessary to reach precisions of the order of 0.01 mag to obtain bounds that could compete with the local ones, but the advantage is that SNIa allows to explore the evolution of the Newton constant with time. Therefore the bounds depend on the cosmological scenario. In the case of the currently favored scenario ($\Omega_M = 0.3$, $\Omega_\Lambda = 0.7$) the bound is $-14 \lesssim \dot{G}_N/G_N \lesssim 26$, while in the case of a flat universe ($\Omega_M = 1.0$, $\Omega_\Lambda = 0.0$) the bound is $-29 \lesssim \dot{G}_N/G_N \lesssim -0.3$ (Lorén-Aguilar et al., 2003).

One of the caveats of this method is the assumption that the fraction of the Chandrasekhar's mass converted into $^{56}Ni$ and the intrinsic luminosity of the supernova is independent of $G_N$, which obviously is not correct. Nevertheless, if $d_L$ can be independently determined, for instance via gravitational-wave standard sirens (Zhao et al., 2018; Zhao and Santos, 2019), it would be possible to trace the evolution of $G_N$ with the redshift.

## 6 NEUTRINO MAGNETIC MOMENTUM

In spite of the enormous progress experienced by the physics of neutrinos, several questions still remain. For instance, are neutrino Dirac or Majorana particles, do sterile neutrinos exist, which is their mass spectrum, do they have magnetic momentum? This last problem, for instance, is specially important since the existence of a magnetic dipole momentum (NMM) can notably enhance the neutrino losses in stars and, consequently, modify the expected evolution.

The magnetic moment couple neutrinos to photons through the effective Lagrangian term:

$$L = -\frac{1}{2}\mu_\nu^{ij}\bar{\psi}_i\sigma_{\alpha\beta}\psi_j F^{\alpha\beta} \quad (17)$$

where $\psi$ is the neutrino field, $F$ the electromagnetic field tensor, $\alpha$ and $\beta$ are Lorentz indices, and $i, j$ the flavor indices. In the SM this interaction is non-zero if neutrinos have non zero mass but is very small due to the specific nature of the SM interactions, where W only couples to left handed currents and the induced value of the magnetic momentum is very small, $\mu_\nu/\mu_B \simeq 3 \times 10^{-19} (m_\nu/1$ eV), where $\mu_B = e/2m_e$ is the Bohr magneton (Marciano and Sanda, 1977). However, in the extensions of the SM this constrain does not apply and $\mu_\nu$ can be large, but not arbitrarily large.

The first experimental bound on the magnetic moment of the electron neutrino was obtained by Cowan and Reines (1957) analyzing the electron recoil spectra in (anti)–neutrino scattering. The GEMMA experiment, specially devoted to detect a hypothetical NMM, obtained is $\mu_\nu < 2.9, \times, 10^{-11}\mu_B$ (Beda et al., 2013).

**Table 4** shows the different bounds obtained up now. Stars, however, can provide better constraints. The first one to use the lifetime of the Sun to constrain the magnetic momentum of the neutrino was Bernstein et al., 1963 who obtained $\mu_\nu \lesssim 10^{-10}\mu_B$. A more recent bound adopting helioseismology argument has been obtained by (Raffelt, 1999) who obtained $\mu_\nu < 4 \times 10^{-10}\mu_B$ and recently the XENON1T detector has provided a bound of $\mu_\nu < 2.9, \times, 10^{-11}\mu_B$ that is comparable to the laboratory values [Aprile et al., 2020].

In the case of massive stars, $8 \lesssim M/M_\odot \lesssim 20$, energy losses by standard neutrinos start to be dominant only when He is almost exhausted in the center of stars and become dominant after carbon ignition. Pair process is the dominant mechanism for the more massive ones while in less massive ones this role is shared by plasma and pair mechanisms. Bremsstrahlung is only important in relatively cool dense stars. If neutrinos have a magnetic momentum, the regions of the $\rho$-$T$ plane in which the different processes are dominant is substantially modified and the evolutionary timescales of stars are strongly perturbed and new effects appear like an increase in the production of C/O white dwarfs and a decrease of the O/Ne WDs and core collapse supernovae (Heger et al., 2009). The bound obtained in this way is $\mu_\nu < 2$–$4 \times 10^{-11}\mu_B$.

After exhausting their hydrogen content in the central layers, low and intermediate mass star evolve along the so-called Red Giant Branch (RGB) in the HR-diagram. These stars are characterized by a He-core surrounded by a H-burning shell. The homology rules show there is a tight relationship between the mass of the helium core and the luminosity and temperature of the burning shell that is almost independent of the mass and metallicity of these stars (Refsdal and Weigert, 1970). Because of the extreme dependence on the temperature, He ignites around $10^8$ K. When this happens stars make an abrupt transition from the RGB to the *red clump* or to the Horizontal Branch (HB) regions, depending on the metallicity, introducing a characteristic discontinuity in the RGB, the so called Tip of the Red Giant Branch (TRGB). Since stars in the TRGB can be easily identified and have a characteristic luminosity they can be used as standard candles. Since this luminosity depends on the size of the core at which helium is ignited (Serenelli et al., 2017), the introduction of any new cooling effect like NMM (or axions) will modify it and the comparison between the empirical calibration and the predictions of the models will provide useful constraints to the new physics input. Applying this method to NGC 4258, a galaxy with a well determined distance, Capozzi and Raffelt (2020) have found a very tight bound, $\mu_\nu < 1.5 \times 10^{-12} \mu_B$[10].

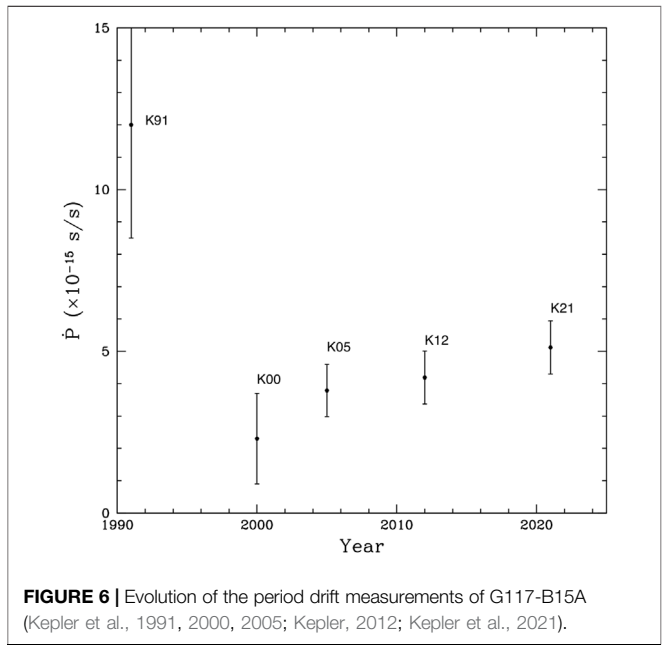

**FIGURE 6** | Evolution of the period drift measurements of G117-B15A (Kepler et al., 1991, 2000, 2005; Kepler, 2012; Kepler et al., 2021).

The importance of neutrino losses in the evolution of white dwarfs was early recognized by Vila (1968) and Savedoff et al., (1969). As it is seen in **Figure 3**, during the first stages of cooling, when the star is still very hot, the energy losses are dominated by neutrinos being the plasma neutrino one dominant $(\gamma \rightarrow \nu\bar{\nu})$ (Iben and Tutukov, 1984; D'Antona and Mazzitelli, 1989). If neutrinos have a magnetic momentum would they could directly couple with photons and the efficiency of the plasma emission would increase. The period drift of DAVs would not be strongly modified by this fact since they are cool but that of DBVs yes[11].

The only reliable estimation of the secular drift of a DBV star is that of PG1351 + 489 (see **Section 4**). Despite the fact that neutrinos are no longer dominant in this star, the existence of an extra cooling due to NMM could be relevant. The analysis performed by Córsico et al., 2014 provided a limit $\mu_\nu \lesssim 7 \times 10^{-12} \mu_B$ that reinforces the values obtained with the TRGB.

As it has been stated in **Sections 2** and **Sections 3**, the analysis based on the early WDLF is affected by the relatively small number of bright white dwarfs and by the uncertainties associated to the initial temperature and chemical profiles of them. The first bound obtained in this way, using the data obtained by Fleming et al., (1986), was $\mu_\nu \lesssim 10^{-11} \mu_B$ (Blinnikov and Dunina-Barkovskaya, 1994). This bound has been improved by Miller Bertolami (2014) using the more precise early luminosity function of Krzesinski et al., (2009) and Rowell and Hambly (2011) and the *ab initio* models of white dwarf computed by Renedo et al., (2010) obtaining $\mu_\nu \lesssim 5 \times 10^{-12} \mu_B$. The analysis of the hot WDLF in 47 Tuc by Hansen et al., (2015) let to obtain $\mu_\nu \lesssim 3.4 \times 10^{-12} \mu_B$

## 7 AXIONS

As it has been mentioned in the Introduction, the SM has been extremely successful but there are still several long standing problems that make the situation highly unsatisfactory. One of them is why the strong interactions do not violate the CP-symmetry despite the presence in the Quantum Chromodynamic Lagrangian of a term that violates this symmetry?

To solve this problem, Peccei and Quinn (1977) introduced a new U (1) symmetry into the Lagrangian of the fundamental interactions that is spontaneously broken at a high energy scale $f_a$. This symmetry implies the existence of a new field *a* which gives raise to a Nambu–Goldstone field that is observed as a new particle, the *axion* (Weinberg, 1978; Wilczek, 1978)[12]. The energy scale is related to the mass of the axion through:

$$m_a \simeq 5.7 \text{ meV} \frac{10^9 \text{GeV}}{f_a} \qquad (18)$$

The mass of axions associated to this energy scale is not fixed by the theory but it has to be smaller enough to ensure a coupling between axions and matter weak enough to account for the lack of detection up to now. Astrophysical and cosmological arguments limit this mass to the range $10^{-6} \text{eV} \lesssim m_a \lesssim 10^{-2} \text{eV}$ (see Turner (1990) and Raffelt (1996) for a complete discussion of such bounds).

Axions couple to photons, electrons and nucleons with strengths that depend on the specific implementation of the Peccei-Quinn symmetry. For instance, in the KSVZ–or hadronic model–axions couple to hadrons and photons only (Kim, 1979; Shifman et al., 1980), while in the DFSZ or GUT model (Dine et al., 1981; Zhitnitskii, 1980), they also couple to charged leptons. Stars can produce axions in analogy with thermal neutrinos and since they can freely escape they are an effective sink of energy. If their mass, not fixed by the theory, is large enough, they can noticeably modify the evolution of stars and reveal their existence. In particular, in the case of white dwarfs, axions can modify the cooling rate as neutrinos do.

---

[10]In previous works, Viaux et al., [2013a,b] obtained $\mu_\nu \lesssim 4.5 \times 10^{-12} \mu_B$ using the red giants of the globular cluster M5.
[11]Notice that the evolution of the period of DOVs would be perturbed but the importance of the radial contraction and the uncertainties of their structure prevent, for the moment, to use them as laboratories.

[12]See Irastorza (2021) for an extensive review about axions and their detection.

The interaction with photons is characterized by the coupling constant $g_{a\gamma}$

$$L_{a\gamma} = -\frac{1}{4}g_{a\gamma}aF_{\mu\nu}\tilde{F}^{\mu\nu}; \quad g_{a\gamma} = \frac{\alpha C_{a\gamma}}{2\pi}\frac{1}{f_a} \quad (19)$$

where $F_{\mu\nu}$ and $a$ represent the electromagnetic field and the axion fields, $\alpha$ the fine structure constant, and $C_{a\gamma}$ a dimensionless coeficient that depends on the axion model.

In the case of the Sun, axions are mainly produced via Primakoff effect and their influence would have an impact on the measured neutrino flux and on the helioseismological observations: The bound that has been obtained is $g_{a\gamma} < 4.1 \times 10^{-10}$ GeV$^{-1}$ (Vinyoles et al., 2015). See **Table 5**.

If the DFSZ implementation is adopted, the interaction of axions with electrons is controlled by the coupling constant $g_{ae}$

$$L_{ae} = -ig_{ae}a\bar{e}\gamma_5 e; \quad g_{ae} = \frac{C_{ae}m_e}{f_a} = 2.8 \times 10^{-14}m_a \cos^2\beta \quad (20)$$

where $C_{ae}$ is a dimensionless coefficient, $m_e$ the mass of the electron and $\tan\beta$ is the ratio between the two Higgs-field expectation values and $m_a$ is the mass of the axion in meV.

As in the case of neutrinos, the tip of the RGB can provide a bound to the interaction electron-axion since these particles are copiously produced in the red giant core (Viaux et al., 2013a). **Table 5** displays the bounds provided by NGC 4258 and $\omega$ Cen. A similar bound was obtained by Straniero et al., (2020) but they also found a positive hint of existence with. $g_e = 0.6^{+0.32}_{-0.58} \times 10^{-13}$

The HB stars are the descendants of RGB stars and, consequently, the ratio between their number $N_{HB}$ and the number of RGB stars brighter than the HB level, $N_{RGB}$ known as the R-parameter, is a measure of the time spent by stars in these regions of the HR-diagram (Iben, 1968). Since HB-stars emit preferentially via Primakoff effect ($\gamma + Ze \rightarrow Ze + a$) the parameter R allows to constrain $g_{a\gamma}$ and $g_{ae}$. The value of R, however, depends on the adopted fraction of helium, Y, for globular clusters and on the existence of rotation (Cassisi et al., 2003). If rotation is neglected, which is a reasonable approach in the case of low-mass stars, Ayala et al., (2014) and Straniero et al., (2015) obtain a bound of $g_{a\gamma} < 0.65 \times 10^{10}$ GeV and a hint of $g_{a\gamma} = 0.3 \pm 0.2 \times 10^{10}$ eV. See **Table 5**.

The hypothesis that an unexpected extra cooling induced by the emission of DFSZ axions was acting in G117-B15A was introduced by Isern et al., (1992), which estimated from **Eq. 13** a value $g_{e,13} \approx 2.4$. Later on this crude estimation was refined using fully evolutionary models (Córsico et al., 2001; Bischoff-Kim et al., 2008; Córsico et al., 2012a; Kepler et al., 2021) and better measurements of the period drift (see **Figure 4**) to obtain $g_{e,13} = 5.66 \pm 0.57$ with the 215 s period. Similar analysis were performed with R548 and L 19–2 (both DAVs) and PG 1351 + 489 (a DBV star) obtaining concordant results (**Table 5**). Notice that the last one has still a strong neutrino emission and is affected by different systematic errors. In particular by an extra emission if neutrinos have a magnetic momentum.

The existence of axions would have two effects on the white dwarf cooling. When the star is hot, if axions are strongly coupled to electrons they modify the thermal profile and reduce the neutrino emission (Miller Bertolami, 2014). When the star cools down neutrino emission is dominated by bremsstrahlung and $\dot{\epsilon}_\nu \propto T^7$ (Itoh and Kohyama, 1983). However, since axions are bossons their bremstrahlung emission behaves as $\dot{\epsilon}_\nu \propto T^4$ (Nakagawa et al., 1987, 1988; Carenza and Lucente, 2021) **Figure 3**, thus modifying the slope of the region where there is the transition from neutrino to photon cooling (Isern et al., 2008), i.e. $8 \lesssim M_{bol} \lesssim 13$. The analysisis of the WDLFs from the SCSS and SDSS catalogues have provided a bound $g_{ae} \leq 2.1 \times 10^{-13}$ and a hint of $g_{ae} = (1.4 \pm 0.3) \times 10^{-13}$ (Miller Bertolami, 2014).

An argument in favor of axion hypothesis comes from the fact that the luminosity functions of the thin and thick discs and halo suggest a shortage of white dwarfs in the luminosity interval $8 \lesssim M_{bol} \lesssim 13$ pointing towards an intrinsic origin. The existence of axions able to interact with electrons with a coupling constant $1.1 \lesssim g_{ae} \times 10^{13} \lesssim 4.5$ is a possibility (Isern et al., 2018), although other ones related with the unsolved DA/non-DA evolution, for instance, can be envisaged. On the contrary, the analysis of the globular cluster 47 Tuc (Hansen et al., 2015; Goldsbury et al., 2016) seems not to favor the axion option although the influence of the presence of a central black hole, the hydrogen burning in low metallicity white dwarfs and the existence of several generations of stars has to be elucidated.

## 8 WEAKLY INTERACTING MASSIVE PARTICLES

The astronomical and cosmological observations indicate that, besides the ordinary baryonic matter, the Universe could contain another component that interacts gravitationally with ordinary matter but it is unable to emit or absorb electromagnetic radiation. It is the so called dark matter (DM). One possibility is provided by WIMPs (Weakly Interacting Massive Particles) that appear in a natural way in some theories trying to go beyond the SM. In order to reach thermal equilibrium in the early Universe and to match the observed DM density they need to have a large mass, $\gtrsim$ 1–100 keV, and self-annihilate with a cross-section $\sigma v \sim 10^{-26}$ cm$^3$s$^{-1}$, where $v$ is the relative velocity between the annihilating particles. These constraints are satisfied by many particles with a mass in the range of MeV-TeV and interactions mediated the exchange of electroweak-scale particles[13] (Bertone and Hooper, 2018).

---

[13]For instance, the neutralino, $\chi_0$, predicted in the context of Supersymmetric theories, is a typical example (Bertone et al., 2005).

At present, the most stringent direct detection bound has been provided by XENON1t (Aprile et al., 2018). In the next future it is expected that XENONnT with a fiducial 4 tones of xenon will provide a bound of $1.4 \times 10^{-48}$ cm$^2$ for a spin-independent WIMP-nucleon interaction and a particle mass of 50 GeV/c$^2$. In the case of spin-dependent interactions this bound would be $2.2 \times 10^{-43}$ cm$^2$ and $6.0, \times, 10^{-42}$ cm$^2$ for interactions with protons and neutrons respectively (Aprile et al., 2020), which are one order of magnitude better than those obtained by XENON1t.

Attempts to obtain information from the gamma-rays, protons-antiprotons or neutrinos-antineutrinos produced by the annihilation or the decay of DM-particles present in the halo, the Galactic Center or dwarf spheroidal galaxies have not produced positive results (Bertone and Hooper, 2018).

If these particles are able to interact, elastically or inelastically, with nucleons they can be captured by stars and planets, settle in their core and perturb their normal evolution (Salati and Silk, 1989; Moskalenko and Wai, 2007; Bertone and Fairbairn, 2008). Neutron stars and white dwarfs are at the focus of such studies because their evolution is just a cooling process and the capture of WIMPs can be an important source of energy via kinetic thermalization and annihilation. In this sense, the analysis of the perturbation of the normal evolution introduced by these particles is probably more promising in the case of white dwarfs than in the case of neutron stars as their structure and evolution is better understood and the observational background more solid.

Dasgupta et al., [2019, 2020]; Bell et al., (2021) have studied in detail the process of capture of DM particles by a white dwarf and their interaction with nucleons and electrons under different hypothesis. In particular, Bell applied these results to white dwarfs present in Messier 4, a globular cluster, using the same hypothesis as McCullough and Fairbairn (2010) concerning the DM properties and have shown that, in the case of DM-nucleon scattering, white dwarfs can probe the sub-GeV region which is not accessible to direct detection searches and, at the same time, can provide competitive values in the $1-10^4$ GeV range. Naturally, these results depend on the DM density in the globular cluster and on a good estimation of its age (Salaris and Cassisi, 2018).

Interestingly enough, Niu et al., (2018) have considered the presence of a DAV star with the same characteristics as G117-15B in the cluster $\omega$-Cen assuming the DM properties proposed by Amaro-Seoane et al., (2016) and have found that the secular drift of the pulsations could provide interesting bounds to the WIMP properties.

## 9 CONCLUSION

White dwarfs can provide useful constraints and hints on many physical speculations. As it has been mentioned there are two ways for doing that, one is through the secular drift of the pulsation period of those that are variables and another is through the mass and luminosity distributions of the Galactic white dwarf populations. Both methods demand models as accurate and precise as possible and that, in turn, demands a deep understanding of the DA, non-DA behaviour, of the equation of state and associated functions, like conductivity, of multicomponent and partially degenerate classical and quantum Coulomb plasmas, as well as convection under partial degeneracy conditions. Concerning variable white dwarfs, besides improving modelling, the main goal should be obtaining reliable period drift s of a representative sample of stars and, concerning mass and luminosity distributions the main problem is obtaining reliable constraints on the IFMR and SFR as well as on galactic evolution, in order to break the existing degeneracy between physical and galactic properties, for which reason **obtaining a** statistically significant sample of the Galactic population of white dwarfs is a must.

## AUTHOR CONTRIBUTIONS

All authors listed have made a substantial, direct, and intellectual contribution to the work and approved it for publication.

## FUNDING

This work has been funded by the Spanish Ministry of Science and Innovation and FEDER UE (MCI-AEI-FEDER, UE) through grants PID2019-108709GB-I00 and the program Unidad de Excelencia María de Maeztu CEX2020-001058-M (JI), by the MINECO grant AYA\-2017-86274-P, and the grant RyC-2016-20254 (AR-M), by grant 2014 SGR 1458 and CERCA Programe of the Generalitat de Catalunya (JI), and by a URICI-CSIC grant.

## ACKNOWLEDGMENTS

We acknowledge the comments and suggestions of P. Lorén-Aguilar, M. Miller-Bertolami and A. Serenelli.